\documentclass[runningheads,a4paper]{llncs}
\usepackage[utf8]{inputenc}
\usepackage{graphicx}
\usepackage{amssymb,graphicx}
\usepackage{xcolor}
\usepackage{dirtytalk}
\usepackage{blindtext}
\usepackage{fancyvrb}
\usepackage{parskip}
\usepackage{wrapfig}
\usepackage[rawfloats=true]{floatrow} 

\usepackage[belowskip=-15pt,aboveskip=0pt]{caption}

\restylefloat{figure} 
\usepackage{hyperref} 
            \hypersetup{backref=true,       
                    pagebackref=true,               
                    hyperindex=true,                
                    colorlinks=true,                
                    breaklinks=true,                
                    urlcolor= black,                
                    linkcolor= blue,                
                    bookmarks=true,                 
                    bookmarksopen=false,
                    citecolor=black,
                    linkcolor=black,
                    filecolor=black,
                    citecolor=blue,
                    linkbordercolor=blue
}
\usepackage{array}
\usepackage{amsmath}
\usepackage{xpatch}
\definecolor{ocre}{RGB}{243,102,25}
\usepackage[labelfont={color=ocre,bf}]{caption}

\newcommand{\bDiamond}{\mathbin{\Diamond}}

\begin{document}

\title{Solving social dilemmas by reasoning about expectations}

\author{Abira Sengupta\inst{1} \and Stephen Cranefield\inst{1} \and Jeremy Pitt\inst{2}
}

\authorrunning{}

\institute{University of Otago, Dunedin, New Zealand \and Imperial College London\\
\email{abira.sengupta@postgrad.otago.ac.nz}, \email{stephen.cranefield@otago.ac.nz},
\email{j.pitt@imperial.ac.uk}}
\maketitle              

\begin{abstract}
It has been argued that one role of social constructs, such as institutions, trust and norms, is to coordinate the expectations of autonomous entities in order to resolve collective action situations (such as collective risk dilemmas) through the coordination of behaviour. While much work has addressed the formal representation of these social constructs, in this paper we focus specifically on the formal representation of, and associated reasoning with, the expectations themselves. In particular, we investigate how explicit reasoning about expectations can be used to encode both traditional game theory solution concepts and social mechanisms for the social dilemma situation. We use the Collective Action Simulation Platform (CASP) to model a collective risk dilemma based on a flood plain scenario and show how using expectations in the reasoning mechanisms of the agents making decisions supports the choice of cooperative behaviour.

\keywords{Collective Action \and Social dilemmas \and Event Calculus \and Expectations.}
\end{abstract}

\section{Introduction}

Collective action takes place when a group of people come together to act in certain ways that will benefit the group as a whole \cite{lee2018collective}.
People have different characteristics, some are more willing to achieve collective benefits rather than their own.
In addition, social norms \cite{ostrom2000collective}, reputation, trust and reciprocity motivate individuals to cooperate \cite{ostrom2010analyzing}. 
Collective action often fails when individuals are not interested in cooperating or coordinating with each other, because there is a conflict between self-interest and collective interest. Thus, individuals pursue their own interests rather than long-term cooperation \cite{olson1965theory} and  ultimately no benefit of collective action occurs \cite{dawes1980social}. For example, the free-rider problem is a type of social dilemma. A free-rider, is someone, who can access a collective benefit without contributing or incurring any cost \cite{hardin2003free,booth1985free}. 

Collective action or social dilemmas are well-known problems, which have been analysed in many studies using the tools of game theory.
Simple game theory models of social dilemmas predict that cooperation is not rational.
However, human society suggests that cooperation can occur due to psychological and social motivations such as benevolence and social norms \cite{sep-social-norms}, internal motivations (e.g.~altruism, fairness norms), rational expectations (e.g.~focal points), social choice mechanisms (e.g.~voting and bargaining) \cite{holzinger2003problems}. These are commonly accommodated in game theory models by modifying the payoff matrix, making cooperation rational. Traditional analyses of this kind of problem in game theory rely on solution concepts such as the Nash Equilibrium. This becomes complex to reason about for a large number of agents and does not seem realistic as a method of human reasoning. In particular, there is a lack of consideration of the (bounded) reasoning processes that can lead community members to participate in collective action \cite{ReubenMPhilThesis}.

In this study, we are interested to model some of the social knowledge underlying collective action directly.
In particular, we believe that explicit reasoning about expectations can explain the achievement of cooperation in social dilemmas. There are already some ideas about expectations from behavioural game theory literature, for example, the Confidence game \cite{geanakoplos1989psychological} where payoffs depend on agent's belief  that represent expectations and the plain-plateau scenario (Maybe we need this) \cite{klein1990microfoundations}, where a credible commitments creates an expectation.  

Our aim is to analyze the various forms of social knowledge that can account for why a human can cooperate in social dilemmas. We posit that this knowledge can often be encoded in the form of expectations, and thus wish to investigate the role of social expectations in agent reasoning mechanisms when faced with collective action problems. Here we investigate how agents might explicitly reason about social knowledge to generate cooperative behaviour in social dilemma situations.

Social expectations  can play an important role to allow cooperation to occur \cite{cranefield2019collective}. In this area, Pitt et al.have developed a computational model for self-governance of common-pool resources based on Ostrom's principles \cite{petruzzi2014social}. Ostrom and Ahn \cite{ahn2003foundations} also investigated three forms of social mechanisms that seem to foster cooperation in collective action problems: those forms are trustworthiness, social network connections and the observed conformance.

To investigate the use of reasoning about expectations to resolve a social dilemma, we model the plain-plateau scenario using the Collective Action Simulation Platform (CASP) \cite{cranefield2019collective} and demonstrate how its support for expectation-based reasoning allows cooperation to take place in this scenario.

The structure of this paper is as follows, in Section 2 we describe the plain-plateau scenario, Section 3 highlights the related concepts and platform, Section 4 describes the modelling of the plain-plateau scenario using (CASP), Section 5 describes about the social solutions for the plain-plateau scenario and Section 6 discusses future work and conclude the paper.

\section{The Plain-Plateau Scenario}

Klein \cite{klein1990microfoundations} introduces a scenario that we refer to as the plain-plateau scenario. In this scenario, the objective of the author is to show how it can be rational for a government to restrict its future choices. This scenario models a society where people can choose to live in a river plain, where they can access water easily; otherwise, they can live on a plateau. When living in the river plain there is a risk of flooding. The government's objective is to maximize the collective utility of the citizens. Klein models citizen's utility as a concave (logarithmic) function of their wealth and house value. Therefore in this scenario when the government has full discretionary power (the \say{discretionary-based regime}), then there is common knowledge that it is in the government's best interest to compensate citizens whose houses have flood damaged by taxing citizens living on the plateau. Thus, citizens who live in the flood plain and suffer due to flood damage can expect to be bailed out by the government. Klein shows that this leads to a prisoner's dilemma game between the citizens, where choosing the plateau is cooperation and choosing the plain is defection\footnote{No compensation resulting as there is no one to tax so (plain, plain) really is sub optimal outcome.}.

To avoid the prisoner's dilemma situation, the government can adopt the \say{rule-based regime} where it removes its own discretionary ability to provide compensation. This can be seen as a binding announcement that the government will not bail out any citizens who have flood damage. In this case, the government will have no reason to tax citizens who are living on the plateau. Therefore, this announcement avoids the prisoner's dilemma.

\subsubsection{Sequence of events}

\begin{figure}[h!]
\includegraphics[scale=0.7]{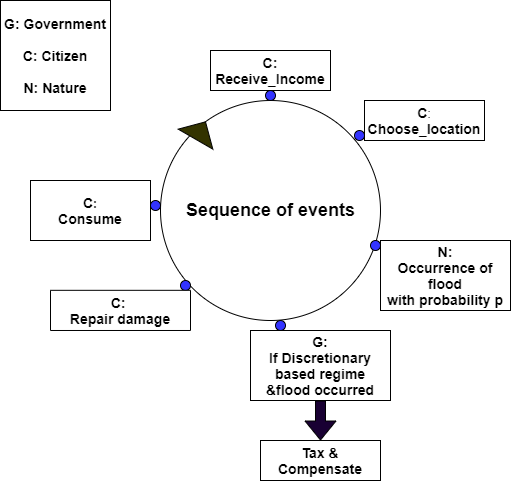}
\caption{Sequence of events in the plain-plateau scenario} \label{fig4}
\end{figure}

Figure 1 shows the occurrence of the events in the plain-plateau scenario. There are multiple rounds where citizens choose where to live. Within each round, the events are as follows:  \,`receive\_income'\, is the step where citizens receive wealth. The next step \,`choose\_location'\, is when agents choose a location to live in (the scenario assumes that their houses can be easily moved). Next there is a  possible flood occurrence with probability $p$. After a flood occurs, if the government has discretionary power then it will tax and compensate. In the next step citizens can repair their houses if they have flood damage. Finally, citizens can consume the remaining money at the \say{consume} step.

Although the plain-plateau scenario is, in general, an n-person game, Klein models this as an extensive form game with two players  \cite{klein1990microfoundations} as illustrated in Figure 2. This scenario involves the government and two citizens, citizen\_1 and citizen\_2 and two action choices, plain and plateau. In this figure, single arrows define the strategies under the rule-based regime and double arrows show strategies under the discretionary-based regime. At state 2 the decision nodes for citizen\_2 form a single information state, i.\,e. citizen\_2 is not aware of what choice was made by citizen\_1. When under the full discretionary power of the government, citizens are better off to live on the plain as the government's rational choice is to compensate for damage. Given Klein’s specific values for periodic income, house values on the plain and plateau, probability of a flood, and amount of flood damage, the government's payoff (social utility) is then 666. 
On the other hand when the government follows the rule-based regime, citizens are better to move to the plateau position and the government's payoff is 730. 

\begin{figure}[h!]
\includegraphics[width =\textwidth]{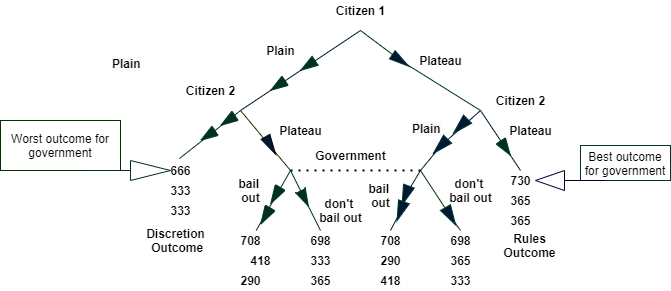}
\caption{Extensive form of plain-plateau game Payoffs are listed in the order: ruler, citizen\_1, citizen\_2.
Single arrows show strategies under the rule-based regime.
Double arrows show strategies under the discretionary-based regime, redraw from [\cite{klein1990microfoundations}].} \label{fig3}
\end{figure}

A Nash equilibrium is \say{a set of strategies, one for each of the n players of a game, that has the property that each player’s choice is his best response to the choices of the n-1 other players} \cite{holt2004nash}. It is interpreted as a potential stable point resulting from agents adjusting their behaviour to search for strategy choices that will give them better results. In particular, a Nash equilibrium is a self-enforcing agreement, which does not need any external power, because due to the self-interest players will follow the agreement if others do.

The prisoner's dilemma game (PD) is a paradox in decision analysis, which shows why two completely rational individuals will not cooperate, even though they would be better off to do so if both of them made that choice \cite{dal2019strategy}. In this game, the highest reward for each party occurs when both players choose to cooperate. However purely rational individuals in the PD game will defect on each other \cite{dal2019strategy}. Figure 3 (a) shows discretionary power of the government leads to a prisoner's dilemma situation with the Nash equilibrium (plain, plain) and payoffs (333, 333) even though there will be no plateau-dweller for the government to tax. On the other hand Figure 3 (b) shows expectation of citizen's under the rule-based regime of the government change the game with different Nash value, (plateau, plateau) and the payoffs (365, 365).

\begin{figure}[h!]
\includegraphics[width =\textwidth]{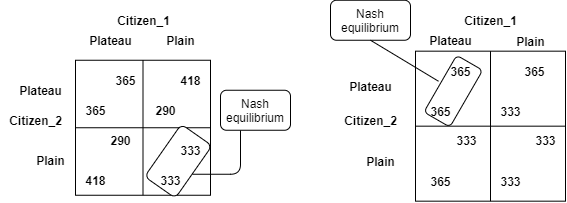}
\caption{(a) The prisoner's dilemma game under the discretionary-based regime and Nash equilibrium (plain, plain) and payoff (333, 333) \cite{klein1990microfoundations}.
    (b) The game under the rule-based regime with Nash equilibrium (plateau, plateau) and payoff (365, 365) }
\end{figure}

\section{Related Concepts and Platform}
The objective of this study is to model and solve the social dilemmas based on the expectations of the agents. This section introduces relevant concepts and tools. We use the CASP platform which is an extension of the Repast Simphony agent-based simulation tool that incorporates an event calculus engine to reason about the physical and social effects of actions.

\subsection{The Event Calculus (EC)}

We use the Event Calculus to model the effects of events in the plain-plateau scenario---both physical (e.g.~damage resulting from flooding) and social (e.g.~the conditional expectation resulting from a government's credible commitment that in the case of flooding no bail-outs will occur).  
The Event Calculus (EC) is a logical language and deductive mechanism to model the effects of actions on information about the state of the world \cite{kowalski1989logic,cranefield2019collective}. Figure 4 illustrates how the event calculus enables a form of reasoning known as temporal projection. This logical language refers to \say{what’s true when given what happens when and what actions do} \cite{shanahan1999event}. The \say{what happened when} is a narrative of events (e.g.~HappensAt(A, T)) and \say{what fluents hold initially} and \say{how events affect fluents} are represented by the vocabulary of the Event Calculus (e.g.~Initially(F), Initiates(A,F,T) and Terminates(A,F,T)).

\begin{figure}[ht!]
\includegraphics[scale=0.5]{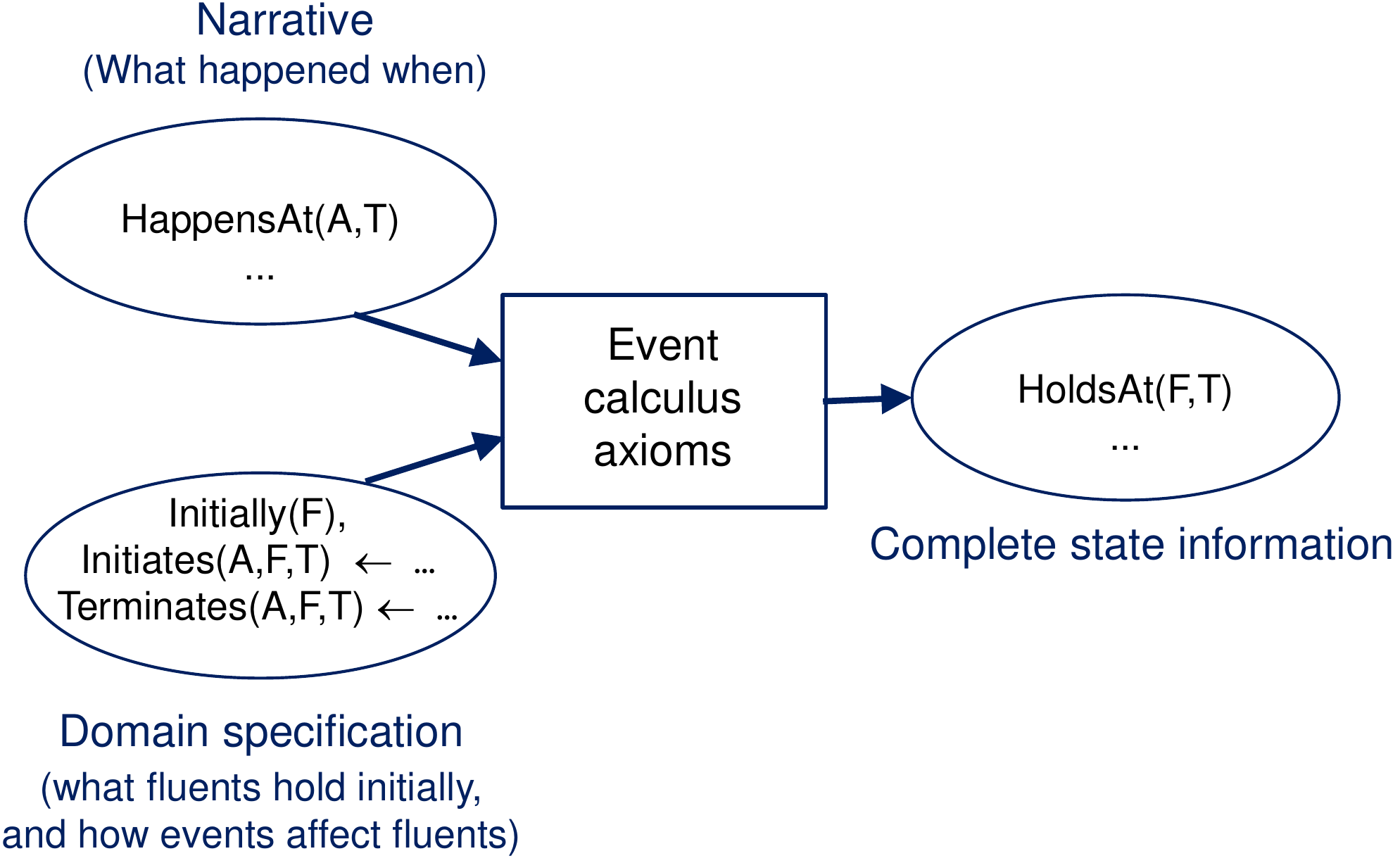}
\caption{Overview of reasoning with the Event Calculus} \label{fig2}
\end{figure}

The Event Calculus models fluents and events. A fluent is a \say{quantity such as \say{the temperature in the room} whose numerical value is subject to variation} \cite{shanahan1999event}. In our work fluents are Boolean e.g.~damage(Agent, Amount). The EC contains an inertia principle which specifies that the value of fluent remains the same until an event occurs that changes the value. 

In this work, we have represented both physical state and social knowledge using a discrete form of the Event Calculus, which has discrete time steps \cite{DBLP:books/mk/Mueller2006}. Within the discrete time points we are able to assign labels to states.
Moreover, the EC is extended with the notion of expectations, and their fulfilment and violation \cite{cranefield2013agents}.

\subsection{Expectations}

An expectation is a future-directed belief that an agent has an active interest in monitoring \cite{castelfranchi2005mind}.
Expectations may be inferred from obligations, commitments, credible announcements, experience, etc. For example, in a circus, a ringmaster may expect his acrobats to form a tower-shaped structure. He needs to know whether his expectation is true or not in future because he has the liability to entertain the audience. In this scenario his belief about his acrobats creates his expectation. 

We model expectations based on the logic of Cranefield \cite{cranefield2013agents}.
We define a fluent $\mathit{exp\_rule} (\mathit{Cond}, \mathit{Exp})$ to state the existence of a conditional rule of expectation: $\mathit{Cond}$ represents a condition of the past and present, and constraint $\mathit{Exp}$ is an expectation regarding the future. If $\mathit{Cond}$ holds then the formula $\mathit{Exp}$ is expected. $\mathit{Cond}$ and $\mathit{Exp}$ are formed from fluents, event occurrences $\mathit{happ}(\mathit{Events})$, Boolean operators, and (for $\mathit{Exp}$ only) Linear Temporal Logic operators e.\,g. $\bigcirc$ (next), $\bDiamond$ (eventually) and $\square$ (always). A fluent of the form $@$L can also be used to test if the current state is associated with the label L. We use labels to model the steps within each cycle in the plain-plateau scenario (e.\,g.~\,`Receive\_Income'\, and \, Change\_location\,').

When the condition of an $\mathit{exp\_rule}$ fluent is true the expectation $\mathit{Exp}$ becomes active, denoted by a fluent $\mathit{exp} (\mathit{Exp})$. An expectation records a state of affairs that is expected to occur, but may not this will be fulfilled or violated if expectation evaluates to true in a state \cite{ranathunga2011integrating}.
$\mathit{Exp}$ may contain temporal operators, so its truth may not be known in the current state. However, it may be able to be partially evaluated. Suppose $\mathit{exp}(p \wedge \bigcirc q)$ holds in the current state (time $t$), i.\,e. $p \wedge \bigcirc q$ is expected where $p$ and $q$ are fluents. If $p$ holds at $t$ then at the next time point $t +1$, the formula $exp(q)$ holds. In other words, $q$ is expected to hold at $t + 1$.

\subsection{The Collective Action Simulation Platform (CASP)}

To simulate agent decision-making with knowledge-based expectations, we use the Collective Action Simulation Platform (CASP), which is an extension of the Java-based Repast Simphony simulation platform \cite{north2013complex}. This allows us to model the effects of actions using a version of the Discrete Event Calculus \cite{DBLP:books/mk/Mueller2006}. It is enhanced by the logic of expectations described above. It also allows agents to take on roles in an institution.

The reasoning of the agent is performed in two stages: (1) a rule engine selects the relevant actions of the agent, which is based on rule related to the agent's current roles. (2) Then the agent can select one of the actions to perform \cite{cranefield2019collective}. Both decisions may involve querying the current state recorded in the event calculus engine. In this work, we have modelled the plain-plateau scenario using CASP.

\section{Modelling the plain-plateau scenario using CASP}

In this section of the paper, we investigate how expectation-based reasoning supports various stages of decision-making in the plain-plateau scenario. Throughout the simulation the citizens do some reasoning to make their decision to perform their actions.

Within the scenario we have two types of agents, \,`Government'\, and \,`Citizen'\, and there are two institutions, \,`Government'\, and \,`Citizens'\,. The citizens' institution has role \,`plain-dweller'\, and \,`plateau-dweller'\, which determine the actions available to the agents in these two locations, plain and plateau.
The government also has institution based roles  \,`discretionary-based regime'\, and \,`rule-based regime'\,.

\begin{center}
\begin{table}
\caption{Roles and actions in the plain-plateau Scenario  }\label{tab1}
\begin{tabular}{|p{2cm}|p{2cm}|p{3.5cm}|p{3.5cm}| }
\hline
Agent &  Institution & Role & Possible Actions\\
\hline
Government &  Government & Rule-based regime & None \\
\hline
Government &  Government  & Discretionary-based regime & tax and compensate\\
\hline
Citizen &  Citizens & Plateau-Dweller & receive income, change role, consume, stay plain\\
\hline
Citizen & Citizens & Plain-Dweller & receive income, change role, repair, consume, stay plateau\\
\hline
\end{tabular}
\end{table}
\end{center}

Information about the agents, their institutions, roles and their possible actions in the simulation are represented in Table 1. 
If the government (be in the \say{government} institution) has the role \,`rule-based regime role'\, then it will perform no action (and specially no tax and compensation). If the government has the \,`discretionary-based regime role'\, then its possible actions are to compensate citizens whose house have been damaged due to flood by taxing citizens who are living on the plateau. Similarly, the institution of the citizen is \,`Citizens'\,. If a citizen (joins the institution \say{citizens}) has the role \,`plateau-dweller'\, then the citizen receives income, then the citizen can stay in that role or change to the \,`plain-dweller'\, role and can stay plain. If citizen changes the role from \,`plateau-dweller'\, to the \,`plain-dweller'\, and due to flood gets damage then repair their damage. Moreover, under this role at the end of a cycle, a citizen can consume the remaining money. On the other hand, when the citizen has the role \,`plain-dweller'\, after joining the institution, the citizen either can stay in that role or change to \,`plateau-dweller'\, role and can stay plateau. The physical effect of changing a villager's role is modelled by event calculus rule that change a \say{location} fluent. With the citizen role, at the beginning citizen receives the income. The citizen can repair his/her damaged house, and in the end, the citizen can consume the remaining money.

Here are examples of the EC clauses used in the simulation:

\begin{verbatim}
initiates(join(government_agent, government, rulesbasedregimerole),
          exp_rule(damage(A,_), not(happ(compensate(A,_)))), _).
\end{verbatim}

This EC clause states that if the government joins the institution `government' with the rule-based regime role then this creates an expectation rule, which defines if a citizen gets damage then there will be no compensation for that citizen.

\begin{verbatim}
initiates(receive_income(A,RMoney), wealth(A,NMoney), T):-
    holdsAt(wealth(A,OMoney), T),
    NMoney is OMoney + RMoney.
\end{verbatim}

This EC clause updates the wealth fluent for an agent when a receive\_income event occurs. The income is added to the existing wealth.
\begin{verbatim}
initiates(change_role(A,citizens, _, citizens_plaindwellerrole),
          location(A,plain), T):-
    holdsAt(member(A,citizens), T).
\end{verbatim}
\begin{verbatim}
initiates(change_role(A, citizens, _, citizens_plateaudwellerrole),
          location(A,plateau), T):-
    holdsAt(member(A,citizens), T).
\end{verbatim}

These EC clauses define the change\_role event for each citizen, who can change his/her role to citizens\_plaindwellerrole if his/her location is plain and if his/her location is plateau then as citizens\_plateaudwellerrole

\begin{verbatim}
initiates(flood, damage(A,D), T):-
    holdsAt(location(A,plain), T),
    \+ holdsAt(damage(A,_), T),
    flood_causes_damage(D).

initiates(flood, damage(A,D), T):-
     holdsAt(damage(A,CD), T),
     initial_house_on_plain_value(V),
     flood_causes_damage(FD),
     D is min(FD + CD, V).
     
terminates(flood, damage(_,_), _).
\end{verbatim}

These EC clauses, the flood event initiates damage for each citizen in the plain. If no damage already exists, a new damage fluent is created. If there is an existing damage fluent, the new fluent records the increased damage, up to the value of the house. Any existing damage fluent is terminated as it is now replaced with an updated one.
\begin{verbatim}
initiates(taxed(A,Tax), wealth(A,New), T) :-
    holdsAt(wealth(A,Old), T),
    New is Old - Tax.
\end{verbatim}

This EC clause updates the wealth fluent for an agent when a taxed event occurs. The tax is deducted from the existing wealth. 

\begin{verbatim}
initiates(compensate(A,Money), wealth(A,New), T):-
   holdsAt(wealth(A,Old), T),
   New is Old + Money.
\end{verbatim}

This EC clause defines the compensate event with two arguments, citizen (A) and money. It initiates a wealth fluent that updates a citizen’s wealth with the amount of compensation.

\begin{verbatim}
initiates(consumed(A), wealth(A,0), _).
\end{verbatim}

This EC clause states that a citizen consumes all the remaining money when the consumed event occurs. The three clauses above have accompanying terminates clauses to remove existing fluents recording damage (for the flood event) and wealth (for taxed and compensate events).

\begin{figure}[h!]
\includegraphics[scale=0.5]{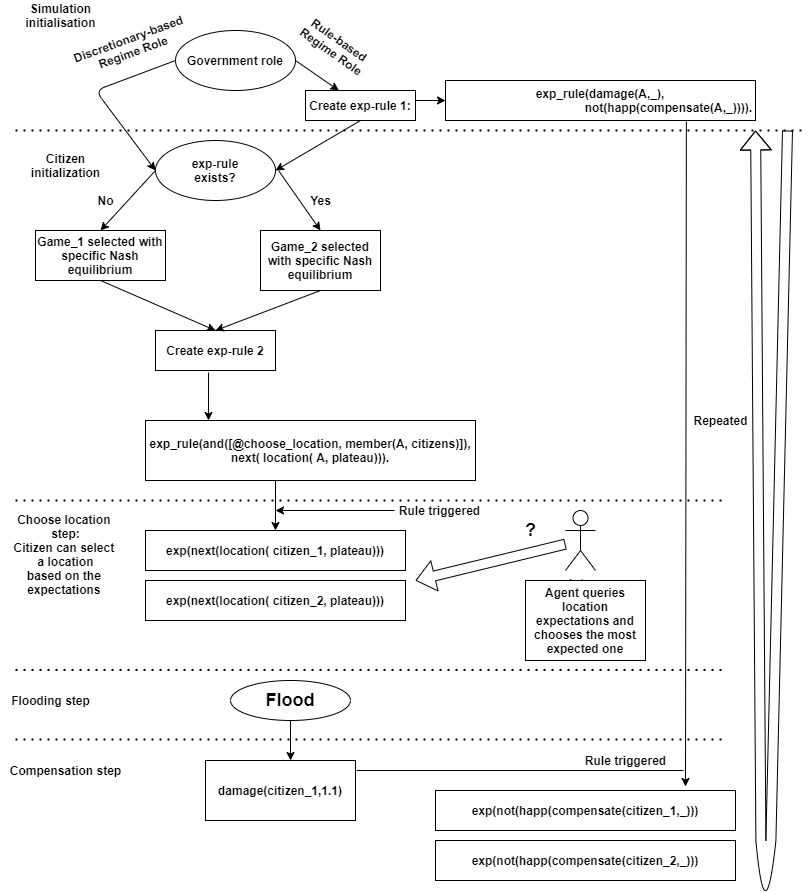}
\caption{Role of expectations of the plain-plateau scenario} \label{fig:fig1}
\end{figure}

Figure 5 shows how expectations are used in the simulation. At the beginning of the simulation (e.g.~simulation initialisation step), the government can join the institution either with the \,`rule-based regime role'\, or the \,`discretionary-based regime role'\,. If the government joins the institution with the  \,`rule-based regime role'\, then an Event Calculus rule initiates an expectation rule for all citizens as shown above is represented as \,`exp\_rule 1'\,. The expectation rule has two arguments, where the first argument is the condition of the rule. The condition of the expectation rule (e.g.~damage(A,\_)) explains that if any citizen gets damage then no compensation will happen for that citizen (e.g.~not(happ(compensate(A,\_)))). On the other hand if the government has full discretionary power then there is no expectation rule generated. When citizens are initiated in the simulation, they check for the existence of this expectation rule. The citizen knows it could be playing one of two possible games. Under the existence of this expectation rule, Game\_2 (Figure 3 (b)) will be selected with the Nash value (plateau, plateau). Otherwise, Game\_1  (Figure 3 (a)) will be selected with the Nash value (plain, plain). The Nash equilibrium\footnote{While we currently use the Nash equilibrium of these games to create these expectations, in principle they could be learned from experience.} is represented as a second exp\_rule fluent, abbreviated `exp\_rule 2'. Here we assume the location is the plateau. After this expectation rule is triggered in the choosing location step, corresponding expectations will be generated for every citizen. Here the \,`exp'\, fluents are created with the expected location of plateau for citizen\_1, and citizen\_2. To make its own location decision, each citizen queries the EC engine for expectations and chooses the most commonly expected location. At the flood step, when the flood event occurs it creates the damage fluent for the citizens who moved to the flood plain. In our scenario, citizen\_1 gets damage after the flooding step. This fluent triggers exp\_rule 1 creating a no compensation expectation for each affected citizen. Any violations of these expectations will be computed by our EC engine and could be used be the agents in this reasoning e.g.~to revise their belief in the government's commitment not to compensate flood victims.

\subsubsection{Evaluation }

Simulating the plain-plateau scenario using CASP confirms that reasoning with expectations successfully generated the appropriate behaviour: cooperation under the rule-based regime and defection under the discretionary-based regime.

This work illustrates how reasoning about expectations using the Discrete Event Calculus and a simulation tool (CASP) allowed us to experiment with mechanisms from game theory and behavioural game theory to explain cooperation resolution of social dilemma using expectations. The explicit representations of expectations were used: 1) to model the government's decision that no compensation would be possible, and 2) the expected location choice of other agents.

\section{Social solutions for the plain-plateau scenario}

Klein notes that the rule-based regime is not the only possible mechanism to promote cooperation in the plain-plateau scenario. In particular, he discusses a range of possible social mechanisms. From Klein’s suggestions, we are interested in experimenting with the use of expectations resulting from social norms to achieve coordination. Suppose that the citizens notice the cost in overall lost utility if anyone chooses to live in the plain, suffers flood damage, and triggers taxation and compensation. A norm may then emerge stating that no one should live in the plain. This norm can be expressed by the following expectation rule, stating that no citizen should move to the plain. 

\noindent\verb|exp_rule(member(A,citizens), never(location(A,plain)))|.

However, this expectation rule may not be enough to encourage the citizens to cooperate. In this situation we may need a second order norm, which states that if the first order norm is violated then any citizen of the institution who moves to the plain can be expected to be punished:

\begin{verbatim}
exp_rule(viol(member(A,citizen),
                   never (location(A,plain)))),
         happ(punish(A)).
\end{verbatim}

We are also interested in reasoning with the expectations underlying team reasoning \cite{lecouteux2018does,sugden2003logic}. In case of expectations based on team reasoning, we intend to use CASP and EC to show how agents are taking their decisions based on the common expectation not to achieve their own individual pay off but instead to achieve the team payoff. Lecouteux~\cite{lecouteux2018does} outlined the following ``team reasoning schema'' (given here for a two-agent team in a social dilemma):
\begin{itemize}
    \item A certain pair of agents are the members of the team.
    \item Each member identifies with the team.
    \item Team members must choose between the joint strategies (C,C), (C,D), (D,C) and (D,D). where C stands for cooperation and D for defection.
    \item The team prefers a strategy that maximises the collection team payoff: (C,C)
    \item Each member wants to choose what the team prefers, so each of us should choose C.
\end{itemize}  
del Corral de Felipe \cite{del_corral_de_felipe_thesis} states some properties of group agency that can be modelled as expectations. For example, membership of a ``collective agent'', e.g.~a team, implies commitment to a certain  \emph{ethos} as a reason for thinking and acting as a team member. In the context of a game theory style interaction defined by a pay-off matrix, that ethos (as expressed in the team reasoning schema above) might be for each member to make decisions that optimise the collective team payoff rather than the individual payoff. This could be expressed by the following expectation rule:
\begin{verbatim}
exp_rule(and([member(Ag, Team, Role), game(Team,G),
              team_optimal(R,G,Act), @action_time]),
         happ(Ag,Act) )
\end{verbatim}

\noindent This states that, for an agent in a team with a specified role, if the team is playing a given game, and it is optimal for an agent in that role playing the game to perform action Act, then once it is time to act, the agent is expected to perform Act.\footnote{This requires a slight extension of our expectation language to allow an actor to be named in a \texttt{happ} term.}

\section{Conclusion and Future work}

In this paper, we have presented an approach to agent-based simulation to show how cooperation towards collective action can be achieved based on reasoning about expectations. To present this investigation we have used the CASP simulation framework, which allows agents to query event calculus fluents representing social knowledge during their reasoning.

Currently, the expectation rule expressing the expected locations of citizens is created at the start of the simulation based on the presence or absence of the government announcement, leading to knowledge of the game being played (Figure 3 (a) or (b)), and then by computing its Nash Equilibrium. In general, we do not plan to rely on the use of Nash Equilibrium. Expectation rules such as this can come from other sources, such as advice from other agents, or learning from observation. 

A feature of our event calculus dialect that has not been used in the plain-plateau simulation so far is the detection of expectation violations and fulfilments. Consider the no-compensation expectations shown at the bottom of Figure~\ref{fig:fig1}. If it turns out that compensation is made to any citizen with flood damage, despite these expectations existing, the Event Calculus engine will create a violation event corresponding to each of these expectations. The agents could choose to monitor for such violations, which may cause them to revise their opinion of the game being played between agents. They could then alter the expectation rule about the location of agents after the next choose\_location step (the plain will now be the rational choice). This will in turn, cause them to choose the plain.

There are several directions for future work. We can find out various solution concepts to solve the collective action problems in a uniform way. This paper discussed expectation-based reasoning in the context of a specific scenario. However, we seek to explore the use of expectation-based reasoning as a general mechanism that can also model other solutions to social dilemmas, e.g. choices influenced by social norms, social capital, team reasoning \cite{lecouteux2018does} etc. In future work, we will investigate the role that expectations play in these social mechanisms to facilitate cooperation in the plain-plateau scenario and in other scenarios.

\bibliographystyle{splncs04}
\bibliography{bibliography}

\begin{thebibliography}{10}
\providecommand{\url}[1]{\texttt{#1}}
\providecommand{\urlprefix}{URL }
\providecommand{\doi}[1]{https://doi.org/#1}

\bibitem{sep-social-norms}
Bicchieri, C., Muldoon, R., Sontuoso, A.: {Social Norms}. In: Zalta, E.N. (ed.)
  The {Stanford} Encyclopedia of Philosophy. Metaphysics Research Lab, Stanford
  University, winter 2018 edn. (2018)

\bibitem{booth1985free}
Booth, A.L.: The free rider problem and a social custom model of trade union
  membership. The Quarterly Journal of Economics  \textbf{100}(1),  253--261
  (1985)

\bibitem{castelfranchi2005mind}
Castelfranchi, C.: Mind as an anticipatory device: For a theory of
  expectations. In: International Symposium on Brain, Vision, and Artificial
  Intelligence. pp. 258--276. Springer (2005)

\bibitem{cranefield2013agents}
Cranefield, S.: Agents and expectations. In: International Workshop on
  Coordination, Organizations, Institutions, and Norms in Agent Systems. pp.
  234--255. Springer (2013)

\bibitem{cranefield2019collective}
Cranefield, S., Clark-Younger, H., Hay, G.: A collective action simulation
  platform. In: International Workshop on Multi-Agent Systems and Agent-Based
  Simulation. pp. 69--80. Springer (2019)

\bibitem{dal2019strategy}
Dal~B{\'o}, P., Fr{\'e}chette, G.R.: Strategy choice in the infinitely repeated
  prisoner's dilemma. American Economic Review  \textbf{109}(11),  3929--52
  (2019)

\bibitem{dawes1980social}
Dawes, R.M.: Social dilemmas. Annual Review of Psychology  \textbf{31}(1),
  169--193 (1980)

\bibitem{del_corral_de_felipe_thesis}
del Corral~de Felipe, M.: The role of commitment in the explanation of agency:
  From practical reasoning to collective action. Ph.D. thesis, National
  University of Distance Education (UNED), Spain (2012)

\bibitem{geanakoplos1989psychological}
Geanakoplos, J., Pearce, D., Stacchetti, E.: Psychological games and sequential
  rationality. Games and Economic Behavior  \textbf{1}(1),  60--79 (1989)

\bibitem{hardin2003free}
Hardin, R., Cullity, G.: The free rider problem. In: The Stanford Encyclopedia
  of Philosophy. E. N. Zalta, Ed. (2013)

\bibitem{holt2004nash}
Holt, C.A., Roth, A.E.: The nash equilibrium: A perspective. Proceedings of the
  National Academy of Sciences  \textbf{101}(12),  3999--4002 (2004)

\bibitem{holzinger2003problems}
Holzinger, K.: The problems of collective action: A new approach. MPI
  Collective Goods Preprint No.{} 2003/2, SSRN (2003), doi:10.2139/ssrn.399140

\bibitem{klein1990microfoundations}
Klein, D.B.: The microfoundations of rules vs. discretion. Constitutional
  Political Economy  \textbf{1}(3),  1--19 (1990)

\bibitem{kowalski1989logic}
Kowalski, R., Sergot, M.: A logic-based calculus of events. In: Foundations of
  knowledge Base Management, pp. 23--55. Springer (1989)

\bibitem{lecouteux2018does}
Lecouteux, G.: What does “we” want? team reasoning, game theory, and
  unselfish behaviours. Revue d'{\'E}conomie Politique  \textbf{128}(3),
  311--332 (2018)

\bibitem{lee2018collective}
Lee, B.H., Struben, J., Bingham, C.B.: Collective action and market formation:
  An integrative framework. Strategic Management Journal  \textbf{39}(1),
  242--266 (2018)

\bibitem{DBLP:books/mk/Mueller2006}
Mueller, E.T.: Commonsense Reasoning. Morgan Kaufmann (2006)

\bibitem{north2013complex}
North, M.J., Collier, N.T., Ozik, J., Tatara, E.R., Macal, C.M., Bragen, M.,
  Sydelko, P.: Complex adaptive systems modeling with repast simphony. Complex
  Adaptive Systems Modeling  \textbf{1}(1), ~3 (2013)

\bibitem{olson1965theory}
Olson, M.: The theory of collective action: public goods and the theory of
  groups. Harvard University Press, Cambridge  (1965)

\bibitem{ostrom2000collective}
Ostrom, E.: Collective action and the evolution of social norms. Journal of
  Economic Perspectives  \textbf{14}(3),  137--158 (2000)

\bibitem{ostrom2010analyzing}
Ostrom, E.: Analyzing collective action. Agricultural Economics  \textbf{41},
  155--166 (2010)

\bibitem{ahn2003foundations}
Ostrom, E., Ahn., T.K.: Foundations of social capital (2003)

\bibitem{petruzzi2014social}
Petruzzi, P.E., Busquets, D., Pitt, J.: Social capital as a complexity
  reduction mechanism for decision making in large scale open systems. In: 2014
  IEEE Eighth International Conference on Self-Adaptive and Self-Organizing
  Systems. pp. 145--150. IEEE (2014)

\bibitem{ranathunga2011integrating}
Ranathunga, S., Purvis, M., Cranefield, S.: Integrating expectation handling
  into jason  (2011)

\bibitem{ReubenMPhilThesis}
Reuben, E.: The Evolution of Theories of Collective Action. {MPhil} thesis,
  Tinbergen Institute (2003)

\bibitem{shanahan1999event}
Shanahan, M.: The event calculus explained. In: Artificial Intelligence Today,
  pp. 409--430. Springer (1999)

\bibitem{sugden2003logic}
Sugden, R.: The logic of team reasoning. Philosophical Explorations
  \textbf{6}(3),  165--181 (2003)

\end{thebibliography}

\end{document}